\documentclass[
reprint,
superscriptaddress,
preprintnumbers,
amsmath,amssymb,
aps,
prb,
]{revtex4-2}

\usepackage[english]{babel}
\usepackage[utf8]{inputenc}
\usepackage[T1]{fontenc}
\usepackage{graphicx}
\usepackage{dcolumn}
\usepackage{bm}
\usepackage{multirow}
\usepackage{hyphenat}
\usepackage[usenames,dvipsnames]{xcolor}

\usepackage{hyperref}
\hypersetup{
final,
colorlinks=true,
linkcolor=blue,
citecolor=blue,
filecolor=blue,
urlcolor=blue}

\usepackage{siunitx}
\DeclareSIUnit{\bohr}{%
\text{bohr}%
}

\usepackage{colortbl}
\usepackage{longtable,booktabs}

\newcommand{\red}[1]{#1}
\renewcommand{\rowcolor}[1]{}

\usepackage{pdfpages} 
\usepackage{pgffor} 
\usepackage{balance}

\makeatletter
\AtBeginDocument{\let\LS@rot\@undefined}
\makeatother

\def\supplementfilename{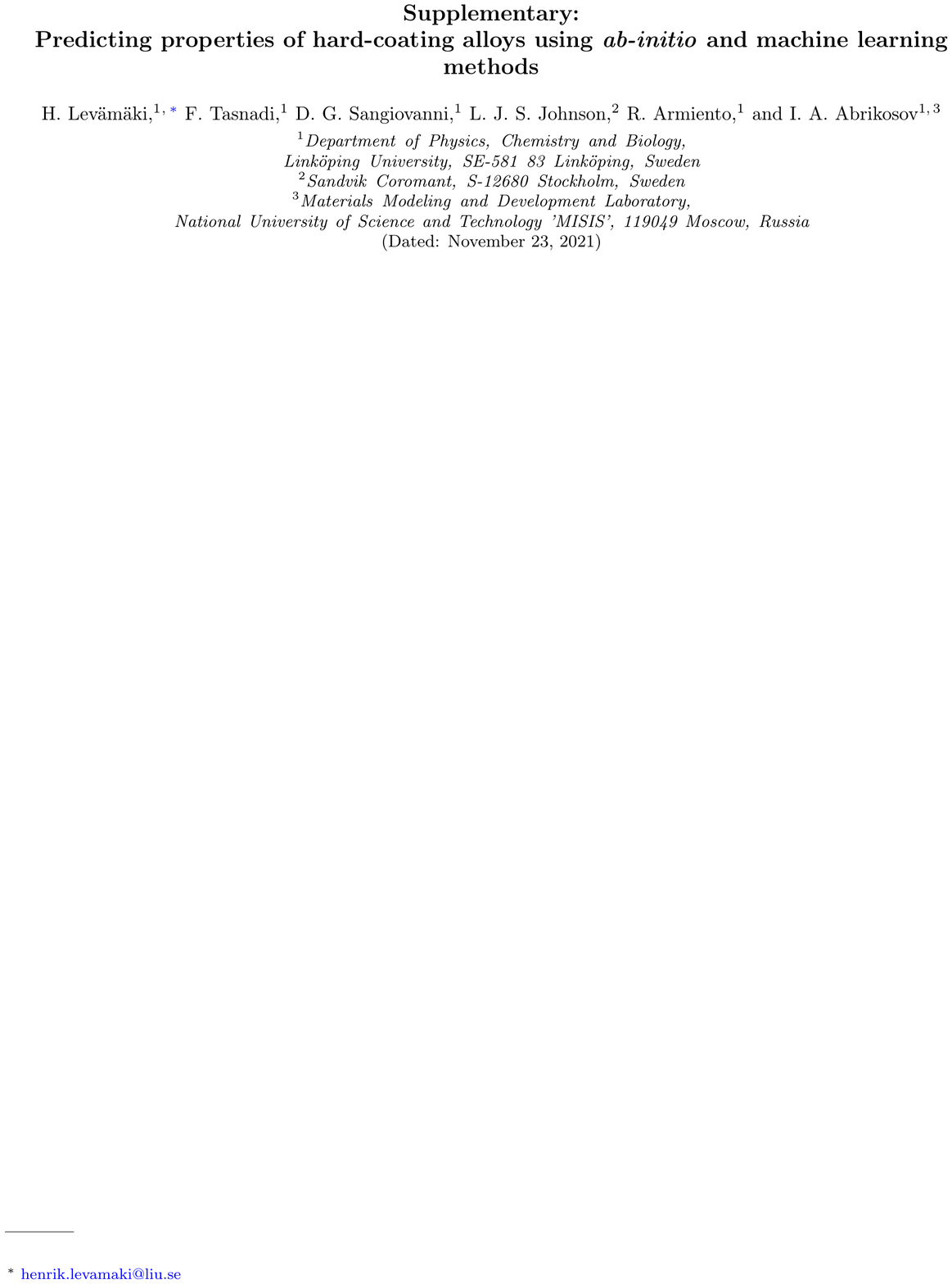}

\pdfximage{\supplementfilename}
\def\numbersupplementpages{\the\pdflastximagepages}

\newif\ifarXiv
\arXivtrue
\begin{document}

\title[Predicting properties of hard-coating alloys using \emph{ab-initio} and machine learning methods]{Predicting properties of hard-coating alloys using \emph{ab-initio} and machine learning methods}%

\author{H. Lev\"am\"aki}%
\email{henrik.levamaki@liu.se}
\affiliation{Department of Physics, Chemistry and Biology, Link\"oping University, SE-581 83 Link\"oping, Sweden}%

\author{F. Tasnadi}
\affiliation{Department of Physics, Chemistry and Biology, Link\"oping University, SE-581 83 Link\"oping, Sweden}%

\author{D. G. Sangiovanni}
\affiliation{Department of Physics, Chemistry and Biology, Link\"oping University, SE-581 83 Link\"oping, Sweden}%

\author{L. J. S. Johnson}
\affiliation{Sandvik Coromant, S-12680 Stockholm, Sweden}

\author{R. Armiento}
\affiliation{Department of Physics, Chemistry and Biology, Link\"oping University, SE-581 83 Link\"oping, Sweden}%

\author{I. A. Abrikosov}
\affiliation{Department of Physics, Chemistry and Biology, Link\"oping University, SE-581 83 Link\"oping, Sweden}%
\affiliation{Materials Modeling and Development Laboratory, National University of Science and Technology 'MISIS', 119049 Moscow, Russia}

\date{\today}

\begin{abstract}
Accelerated design of novel hard coating materials requires state-of-the-art computational tools,  which include data-driven techniques, building databases, and training machine learning models against the databases.
In this work, we present a development of a heavily automated high-throughput workflow to build a database of industrially relevant hard coating materials, such as binary and ternary nitrides.
We use Vienna Ab initio Simulation package as the density functional theory calculator and the high-throughput toolkit to automate the calculation workflow.
We calculate and present results, including the elastic constants, one of the key materials parameter that determines mechanical properties of the coatings, for $X_{1-x}Y_x$N binary and ternary nitrides, where $X$,$Y$ $\in \{\text{Al}, \text{Ti}, \text{Zr}, \text{Hf}\}$ and fraction $x = 0, \frac{1}{4}, \frac{1}{2}, \frac{3}{4}, 1$.
We explore ways for ML techniques to support and complement the designed databases.
We find that the crystal graph convolutional neural network model trained on Materials Project data for ordered lattices has sufficient prediction accuracy for the disordered nitrides, suggesting that the existing databases provide important data for predicting mechanical properties of qualitatively different type of material systems, in our case hard coating alloys, not included in the original dataset.
\end{abstract}



\maketitle

\section{Introduction} \label{sec:introduction}
Hard coating materials have a wide range of applications including metal cutting,  scratch resistant coatings, grinding, and aerospace and automotive parts.
For metal cutting tools, hard coatings are essential to their machining performance, increasing it by orders of magnitude over uncoated tools in most applications.
Since the beginning, high-hardness materials have been developed experimentally, but for a few decades computational tools, such as density functional theory (DFT) \cite{Hohenberg1964,Kohn1965} and molecular dynamics, have been used to support and complement experiments.
However, as the design of new coatings moves from binary and ternary systems to quaternary and beyond, the combinatorial complexity increases rapidly, making conventional approaches increasingly difficult.

More recently, machine learning (ML) has been found to be a viable way to reduce the number of experiments, as well as computations, to accelerate the design process \cite{Bhadeshia1999,Tehrani2018,Mazhnik2020,Avery2019,Bedolla2020}.
Demand for robust ML models is there, but a big hurdle in their adoption is the limited availability of data, either experimental or computational, that ML models need to be trained on.
Experimental data has traditionally been relatively scarce, while computational databases, such as Materials Project \cite{Jain2013} and AFLOW \cite{Curtarolo2012}, have grown to contain tens or hundreds of thousands of entries.
For example, Avery \emph{et al.} used AFLOW and a combined ML and evolutionary search method to predict new superhard phases in carbon \cite{Avery2019}.
Mazhnik \emph{et al.} used Materials Project data to train the crystal graph convolutional neural network (CGCNN) \cite{Xie2018} model to predict the bulk and shear moduli of ordered compounds using only the structural and chemical information of the system as input \cite{Mazhnik2020}, obtaining good results.
\red{
In this paper we extend the approach to qualitatively different class of materials, disordered alloys (as deposited metastable thin films).
It is well-known that ML is good at interpolation, but bad at extrapolation, and it is therefore vital to establish how well ML models extrapolate from ordered data to disordered data.
We utilize a transfer learning (TL) approach \cite{Yamada2019} to predict polycrystalline elastic constants of disordered alloys based on data obtained from ordered compounds.
To our knowledge, there are only very few TL studies in predicting properties of alloys.
For a broad overview of ML approaches utilized in case of alloys, see \cite{hartMachineLearningAlloys2021}.
}

From the point of view of practical applications, e.g., to design new hard coatings, however, currently existing databases are lacking for two reasons.
Firstly, elastic constants are important: for instance, the intrinsic hardness of a material can be qualitatively assessed from the elastic constants and moduli \cite{tseIntrinsicHardnessCrystalline2010,Tehrani2019}.
However, since elastic constants are demanding to calculate, only a limited fraction of the entries in existing databases have elastic constants data included.
Secondly, most industrially relevant hard coatings, e.g., Ti$_{1-x}$Al$_x$N, Cr$_{1-x}$Al$_x$N, and Ti$_{1-x}$Si$_x$N, are substitutionally disordered \cite{Abrikosov2011}, and are often thermodynamically metastable or even unstable, which is possible due to the far-from-equilibrium  synthesis techniques of physical vapor deposition.
Existing databases, on the other hand, are focused on ordered compounds.
Creating databases for disordered systems is therefore an activity that requires more attention.
There is a recent effort to enable data-driven study of high-entropy ceramics \cite{osesHighentropyCeramics2020}, and in this paper we work towards a database of disordered hard coating materials.
Because of significant computational costs, direct first-principles calculations for disordered alloys would lead to a relatively small database, which limits the amount of data to train ML models on.
On the other hand, there are recent efforts in literature to find ways to make ML useful for small datasets \cite{Yamada2019,Breuck2020,Lee2021,Zhang2018c,Chen2021}.

Given the abundance of data for ordered compounds, we investigate how ordered data can help with the lack of disordered data.
Using a bigger data set to help the modeling of a smaller data set is one of the goals of TL \cite{Lee2021}.
In this paper we use a TL approach, in which the CGCNN model is trained on the data for ordered compounds available in Materials Project, and the model is then used to predict the properties of disordered systems.
We are interested in predicting quantities that are known to correlate with the intrinsic hardness of materials, such as bulk and shear moduli \cite{Oganov2010,Tian2012,Kvashnin2019,Mazhnik2019}.
We find that even without an explicit inclusion of the data for disordered alloys in the model, it is able to predict the bulk and shear moduli of the industrially relevant nitrides with sufficient accuracy.

\section{Results} \label{sec:results_discussion}
\subsection{DFT results} \label{subsec:dft_results}
\begin{table}
	\caption{\label{tab:MAE_table_ar} Mean absolute error (MAE) in GPa (except the unitless Poisson's ratio $\nu$) and mean absolute relative error (MARE) $(X_{\text{rel}} - X_{\text{unrel}})/X_{\text{unrel}}$ in $\%$ between the relaxed and unrelaxed elastic properties.}
	\begin{tabular}{crrrrrrrrr}
		\multicolumn{10}{c}{MAE}\\
		\hline
		& $c_{11}$ & $c_{12}$ & $c_{13}$ & $c_{33}$ & $c_{44}$ & $B$ & $G$ & $E$ & $\nu$ \\
		\hline
		B$1$ & 8 & 6 & --- & --- & 9 & 4 & 4 & 10 & 0.004\\
		B$3$ & 4 & 3 & --- & --- & 38 & 3 & 17 & 38 & 0.035\\
		B$4$ & 41 & 34 & 39 & 108 & 11 & 3 & 26 & 57 & 0.052\\
		\hline
		\hline
		&&&&&&&&&\\
		\rowcolor{red}\multicolumn{10}{c}{MARE}\\
		\hline
		\rowcolor{red}& $c_{11}$ & $c_{12}$ & $c_{13}$ & $c_{33}$ & $c_{44}$ & $B$ & $G$ & $E$ & $\nu$ \\
		\hline
		\rowcolor{red}B$1$ & 2 & 5 & --- & --- & 6 & 2 & 3 & 3 & 2\\
		\rowcolor{red}B$3$ & 2 & 2 & --- & --- & 28 & 2 & 17 & 15 & 13\\
		\rowcolor{red}B$4$ & 12 & 31 & 46 & 32 & 14 & 2 & 26 & 23 & 21\\
		\hline
		\hline
	\end{tabular}
\end{table}

\begin{table}
	\footnotesize
	\caption{\label{tab:all_results} Calculated DFT elastic constants, bulk ($B$), shear ($G$) and Young's moduli ($E$) in GPa, as well as Poisson's ratio ($\nu$). The SB column refers to the Strukturbericht designation.
	Those B4 systems that relaxed into the B$_k$ structure are marked by ``B4 $\rightarrow$ B$_k$'' in the SB column.}
	\begin{tabular}{llrrrrrrrrr}
	\hline
	\hline
	Alloy & SB & $c_{11}$ & $c_{12}$ & $c_{13}$ & $c_{33}$ & $c_{44}$ & $B$ & $G$ & $E$ & $\nu$ \\
	\hline
	AlN & B1 & 432 & 167 & --- & --- & 307 & 256 & 219 & 511 & 0.167\\ 
	HfN & B1 & 598 & 112 & --- & --- & 129 & 274 & 166 & 414 & 0.248\\ 
	Hf$_{0.75}$Al$_{0.25}$N & B1 & 524 & 124 & --- & --- & 147 & 258 & 166 & 410 & 0.235\\ 
	Hf$_{0.50}$Al$_{0.50}$N & B1 & 460 & 142 & --- & --- & 174 & 248 & 168 & 411 & 0.224\\ 
	Hf$_{0.25}$Al$_{0.75}$N & B1 & 429 & 158 & --- & --- & 221 & 248 & 182 & 439 & 0.205\\ 
	Hf$_{0.75}$Ti$_{0.25}$N & B1 & 581 & 113 & --- & --- & 134 & 269 & 168 & 417 & 0.242\\ 
	Hf$_{0.50}$Ti$_{0.50}$N & B1 & 567 & 119 & --- & --- & 142 & 268 & 171 & 423 & 0.237\\ 
	Hf$_{0.25}$Ti$_{0.75}$N & B1 & 577 & 122 & --- & --- & 153 & 274 & 180 & 443 & 0.231\\ 
	Hf$_{0.75}$Zr$_{0.25}$N & B1 & 584 & 109 & --- & --- & 128 & 267 & 165 & 410 & 0.244\\ 
	Hf$_{0.50}$Zr$_{0.50}$N & B1 & 569 & 108 & --- & --- & 128 & 262 & 162 & 403 & 0.244\\ 
	Hf$_{0.25}$Zr$_{0.75}$N & B1 & 554 & 107 & --- & --- & 127 & 256 & 159 & 395 & 0.243\\ 
	TiN & B1 & 584 & 136 & --- & --- & 166 & 285 & 187 & 460 & 0.231\\ 
	Ti$_{0.75}$Al$_{0.25}$N & B1 & 535 & 139 & --- & --- & 181 & 271 & 187 & 456 & 0.22\\ 
	Ti$_{0.50}$Al$_{0.50}$N & B1 & 492 & 155 & --- & --- & 214 & 268 & 194 & 469 & 0.208\\ 
	Ti$_{0.25}$Al$_{0.75}$N & B1 & 443 & 161 & --- & --- & 250 & 255 & 199 & 474 & 0.19\\ 
	ZrN & B1 & 544 & 110 & --- & --- & 125 & 255 & 156 & 389 & 0.246\\ 
	Zr$_{0.75}$Al$_{0.25}$N & B1 & 477 & 124 & --- & --- & 139 & 242 & 153 & 379 & 0.239\\ 
	Zr$_{0.50}$Al$_{0.50}$N & B1 & 427 & 141 & --- & --- & 166 & 236 & 156 & 384 & 0.229\\ 
	Zr$_{0.25}$Al$_{0.75}$N & B1 & 411 & 157 & --- & --- & 215 & 242 & 174 & 421 & 0.21\\ 
	Zr$_{0.75}$Ti$_{0.25}$N & B1 & 538 & 111 & --- & --- & 130 & 253 & 159 & 394 & 0.24\\ 
	Zr$_{0.50}$Ti$_{0.50}$N & B1 & 540 & 116 & --- & --- & 138 & 257 & 164 & 406 & 0.237\\ 
	Zr$_{0.25}$Ti$_{0.75}$N & B1 & 568 & 122 & --- & --- & 150 & 271 & 176 & 434 & 0.233\\ 
	&&&&&&&&&\\
	AlN & B3 & 285 & 152 & --- & --- & 178 & 196 & 120 & 299 & 0.246\\ 
	HfN & B3 & 292 & 152 & --- & --- & 93 & 199 & 83 & 219 & 0.317\\ 
	Hf$_{0.75}$Al$_{0.25}$N & B3 & 274 & 144 & --- & --- & 100 & 188 & 84 & 219 & 0.306\\ 
	Hf$_{0.50}$Al$_{0.50}$N & B3 & 261 & 144 & --- & --- & 106 & 183 & 84 & 219 & 0.301\\ 
	Hf$_{0.25}$Al$_{0.75}$N & B3 & 263 & 144 & --- & --- & 127 & 183 & 94 & 241 & 0.281\\ 
	Hf$_{0.75}$Ti$_{0.25}$N & B3 & 287 & 150 & --- & --- & 93 & 196 & 82 & 216 & 0.316\\ 
	Hf$_{0.50}$Ti$_{0.50}$N & B3 & 286 & 150 & --- & --- & 92 & 195 & 82 & 216 & 0.316\\ 
	Hf$_{0.25}$Ti$_{0.75}$N & B3 & 289 & 152 & --- & --- & 92 & 198 & 82 & 216 & 0.318\\ 
	Hf$_{0.75}$Zr$_{0.25}$N & B3 & 281 & 148 & --- & --- & 91 & 193 & 80 & 211 & 0.318\\ 
	Hf$_{0.50}$Zr$_{0.50}$N & B3 & 272 & 147 & --- & --- & 90 & 189 & 78 & 206 & 0.319\\ 
	Hf$_{0.25}$Zr$_{0.75}$N & B3 & 263 & 144 & --- & --- & 88 & 184 & 75 & 198 & 0.321\\ 
	TiN & B3 & 293 & 155 & --- & --- & 91 & 201 & 81 & 214 & 0.322\\ 
	Ti$_{0.75}$Al$_{0.25}$N & B3 & 282 & 154 & --- & --- & 103 & 197 & 85 & 223 & 0.311\\ 
	Ti$_{0.50}$Al$_{0.50}$N & B3 & 274 & 154 & --- & --- & 116 & 194 & 89 & 232 & 0.301\\ 
	Ti$_{0.25}$Al$_{0.75}$N & B3 & 271 & 153 & --- & --- & 138 & 192 & 98 & 251 & 0.282\\ 
	ZrN & B3 & 258 & 144 & --- & --- & 87 & 182 & 73 & 193 & 0.323\\ 
	Zr$_{0.75}$Al$_{0.25}$N & B3 & 247 & 140 & --- & --- & 92 & 176 & 74 & 195 & 0.316\\ 
	Zr$_{0.50}$Al$_{0.50}$N & B3 & 240 & 142 & --- & --- & 98 & 175 & 74 & 195 & 0.315\\ 
	Zr$_{0.25}$Al$_{0.75}$N & B3 & 253 & 146 & --- & --- & 123 & 182 & 88 & 227 & 0.292\\ 
	Zr$_{0.75}$Ti$_{0.25}$N & B3 & 262 & 144 & --- & --- & 87 & 184 & 74 & 196 & 0.323\\ 
	Zr$_{0.50}$Ti$_{0.50}$N & B3 & 269 & 148 & --- & --- & 88 & 188 & 76 & 201 & 0.322\\ 
	Zr$_{0.25}$Ti$_{0.75}$N & B3 & 282 & 154 & --- & --- & 89 & 197 & 78 & 207 & 0.325\\ 
	&&&&&&&&&\\
AlN & B4 & 376 & 129 & 98 & 352 & 112 & 194 & 121  & 301 & 0.242\\ 
HfN & B4 & 253 & 161 & 162 & 220 & 47 & 188 & 44  & 122 & 0.391\\ 
Hf$_{0.75}$Al$_{0.25}$N & B4 $\rightarrow$ B$_k$ & 296 & 204 & 113 & 359 & 107 & 201 & 79  & 210 & 0.326\\ 
Hf$_{0.50}$Al$_{0.50}$N & B4 $\rightarrow$ B$_k$ & 298 & 176 & 145 & 110 & 109 & 135 & 51  & 136 & 0.332\\ 
Hf$_{0.25}$Al$_{0.75}$N & B4 & 305 & 140 & 124 & 216 & 87 & 175 & 79  & 206 & 0.304\\ 
Hf$_{0.75}$Ti$_{0.25}$N & B4 $\rightarrow$ B$_k$ & 298 & 210 & 106 & 493 & 113 & 214 & 85  & 225 & 0.325\\ 
Hf$_{0.50}$Ti$_{0.50}$N & B4 $\rightarrow$ B$_k$ & 304 & 210 & 103 & 491 & 120 & 214 & 90  & 237 & 0.316\\ 
Hf$_{0.25}$Ti$_{0.75}$N & B4 $\rightarrow$ B$_k$ & 311 & 212 & 106 & 500 & 128 & 218 & 94  & 247 & 0.311\\ 
Hf$_{0.75}$Zr$_{0.25}$N & B4 $\rightarrow$ B$_k$ & 292 & 211 & 105 & 495 & 106 & 212 & 81  & 216 & 0.331\\ 
Hf$_{0.50}$Zr$_{0.50}$N & B4 $\rightarrow$ B$_k$ & 288 & 206 & 102 & 483 & 107 & 208 & 81  & 215 & 0.328\\ 
Hf$_{0.25}$Zr$_{0.75}$N & B4 $\rightarrow$ B$_k$ & 286 & 201 & 101 & 470 & 107 & 205 & 82  & 217 & 0.324\\ 
TiN & B4 $\rightarrow$ B$_k$ & 322 & 219 & 111 & 524 & 138 & 227 & 100  & 262 & 0.308\\ 
Ti$_{0.75}$Al$_{0.25}$N & B4 $\rightarrow$ B$_k$ & 311 & 212 & 120 & 438 & 138 & 218 & 93  & 244 & 0.313\\ 
Ti$_{0.50}$Al$_{0.50}$N & B4 $\rightarrow$ B$_k$ & 314 & 196 & 128 & 363 & 148 & 211 & 98  & 255 & 0.299\\ 
Ti$_{0.25}$Al$_{0.75}$N & B4 & 315 & 136 & 136 & 214 & 88 & 181 & 80  & 209 & 0.307\\ 
ZrN & B4 $\rightarrow$ B$_k$ & 284 & 195 & 101 & 458 & 106 & 202 & 82  & 217 & 0.321\\ 
Zr$_{0.75}$Al$_{0.25}$N & B4 $\rightarrow$ B$_k$ & 277 & 195 & 108 & 320 & 101 & 188 & 72  & 192 & 0.33\\ 
Zr$_{0.50}$Al$_{0.50}$N & B4 $\rightarrow$ B$_k$ & 287 & 177 & 125 & 202 & 110 & 176 & 74  & 195 & 0.316\\ 
Zr$_{0.25}$Al$_{0.75}$N & B4 & 302 & 147 & 122 & 190 & 89 & 169 & 76  & 198 & 0.304\\ 
Zr$_{0.75}$Ti$_{0.25}$N & B4 $\rightarrow$ B$_k$ & 288 & 200 & 101 & 453 & 110 & 203 & 83  & 219 & 0.32\\ 
Zr$_{0.50}$Ti$_{0.50}$N & B4 $\rightarrow$ B$_k$ & 297 & 205 & 102 & 461 & 116 & 207 & 87  & 229 & 0.316\\ 
Zr$_{0.25}$Ti$_{0.75}$N & B4 $\rightarrow$ B$_k$ & 307 & 210 & 105 & 485 & 125 & 215 & 92  & 242 & 0.313\\
\hline
	\end{tabular}
\end{table}

\begin{figure}
	\begin{center}
		\includegraphics[width=1.0\linewidth,height=0.8\textheight,keepaspectratio]{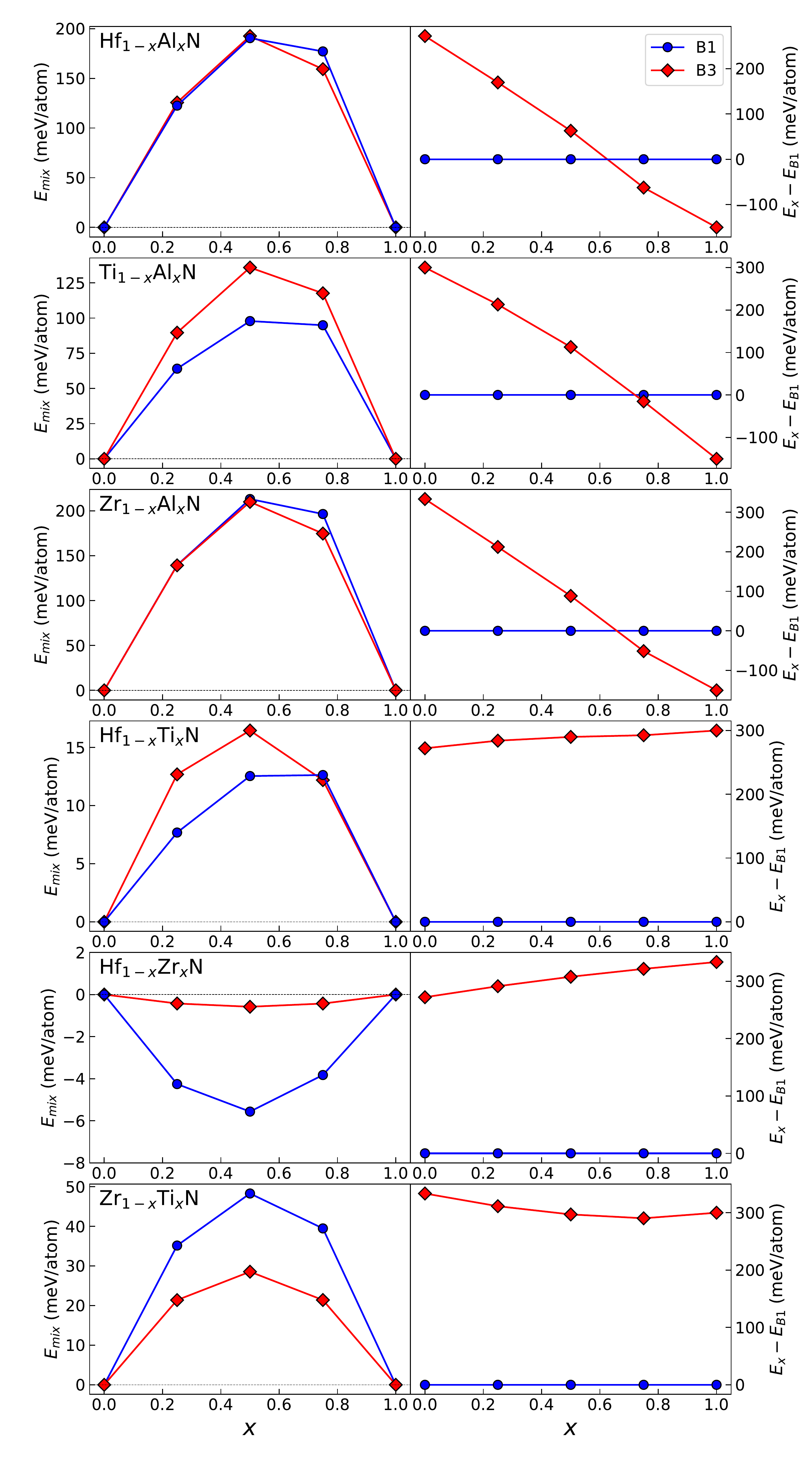}
	\end{center}
	\caption{\label{fig:mixing_energies} Left panel: calculated DFT mixing energies $E[X_{1-x}Y_x\text{N}] - (1-x)E[X\text{N}] - xE[Y\text{N}]$. Right panel: structural stability with respect to B1 with the same composition $E[\text{structure}] - E[\text{B}1]$. B$4$ alloys, most of which relaxed into the B$_k$ setting (see Table~\ref{tab:all_results}) are not shown.}
\end{figure}

We calculate disordered binary and ternary nitrides of the form $X_{1-x}Y_x$N, where $X$,$Y$ $\in \{\text{Al}, \text{Ti}, \text{Zr}, \text{Hf}\}$ and the concentration parameter $x = 0, \frac{1}{4}, \frac{1}{2}, \frac{3}{4}, 1$.
The calculations are performed with and without taking atomic relaxations into account, so that we can assess the importance of the atomic relaxation effects.
Cell shape and volume are optimized in both cases.
Note that in this comparison in the unrelaxed B$4$ calculations we must fix the one available atomic degree of freedom along the $z$-direction.
We fix it by setting the atomic unit cell coordinates as $z_1 = 0$ and $z_2 = 0.38$ in terms of the notation of Ref.~\cite{B4_prototype}.
One can define $z_1 = 0$ without loss of generality and $z_2 = 0.38$ represents the average value for the type of systems we consider here.
It should be noted that the $z_2$ coordinate is also called the ``$u$ parameter'' in some literature sources \cite{kisiParametersWurtziteStructure1989}.

Let us first investigate the effect of atomic relaxations on the elastic constants.
Assessment of the computational cost due to atomic relaxation is useful, because computing the elastic constants is time consuming. Thus, if the effect is minor, some computational time could be saved by neglecting atomic relaxations.
Several previous studies have investigated the impact that atomic relaxations have on the elastic constant results \cite{Tian2015,Tian2016a,Song2017,Tian2017,Kim2019,Jafary_Zadeh_2019,Taga2005a}.
In summary, the previous studies have found that, in most cases, relaxed and unrelaxed elastic constants are quite similar.
Most often, the relative differences are lower than $\approx10\%$.
Relaxation effects can also be qualitatively estimated from symmetry arguments.
For some lattice symmetries, a uniform deformation (used for elastic-constant calculations) may be sufficient to shift atoms from their equilibrium positions.
This implies, in turn, that the resulting elastic tensors strongly depend on the strain matrix used for the calculation.

Table~\ref{tab:MAE_table_ar} shows the mean absolute errors (MAE) and mean absolute relative errors (MARE) $(X_{\text{rel}} - X_{\text{unrel}})/X_{\text{unrel}}$ between the relaxed and unrelaxed elastic constants and the polycrystalline quantities,
\red{calculated for all the alloys that can be found in Table~\ref{tab:all_results}, except for those that are marked ``B4 $\rightarrow$ B$_k$''.}
We have not included those B4 structures that relaxed into B$_k$, because all unrelaxed calculations remain in the B4 phase and therefore are not directly comparable to the relaxed calculations that end up in the B$_k$ phase (see Fig.~\ref{figure3}).
We can see that for the B$1$, which is a highly symmetric structure, relaxation effects are minor.
With the B$3$ phase, atomic relaxation effects on calculated $c_{11}$ and $c_{12}$ are negligible.
Conversely, the MAE and MARE values indicate a strong effect of atomic relaxation on $c_{44}$.
That is due to local B3-symmetry-breaking induced by shear lattice distortions.
Also the B$4$ phase shows noticeable relaxation effects.

Based on our findings, if one is only interested in the bulk modulus, one might perform only unrelaxed calculations.
If the shear modulus, or quantities that depend on the shear modulus, is of interest, the unrelaxed shear modulus will be accurate only for structures such as B$1$, where the different distortions do not break the symmetry in such a way that there are significant distortion dependent atomic movements.

The calculated elastic properties are shown in Table~\ref{tab:all_results} and the mixing energies $E[X_{1-x}Y_x\text{N}] - (1-x)E[X\text{N}] - xE[Y\text{N}]$ and relative stabilities $E[\text{structure}] - E[\text{B}1]$ of the B1 and B3 structure types are shown in Fig.~\ref{fig:mixing_energies}.
\red{The B$4$ alloys, most of which relaxed into the B$_k$ setting are not shown.}
All of the calculated alloys are found to be mechanically stable within a $10\%$ tolerance of $\epsilon=1.1$ (see Sec. \ref{subsec:dft_methods}).
Dynamical stability checks, which are based on phonon dispersion calculations, are out of the scope of this work, but based on existing literature some of the calculated alloys can be expected to be dynamically unstable.
All two-component systems are ordered and those systems have been checked against available Materials Project data and the two sets have been found to be generally in good agreement.
Our results for the well-studied Ti$_{1-x}$Al$_x$N system are also in agreement with literature data \cite{Lind2011,Abrikosov2011,Tasnadi2012,Shulumba2015}.

For the alloys in B1 structure we find good agreement with the elastic constants, bulk modulus, and mixing energy results of Refs.~\cite{Holec2013,Wang2017}.
\red{Ref.~\cite{Holec2013} calculated the same set of alloys that can be found in Fig.~\ref{fig:mixing_energies}, except for Hf$_{1-x}$Zr$_x$N.
The mixing energies of Fig.~\ref{fig:mixing_energies} are in general in good agreement with those of Ref.~\cite{Holec2013}, except for the trend that Fig.~\ref{fig:mixing_energies} obtained mixing energies that are highly symmetrical in terms of the concentration parameter $x$ axis.
Our mixing energies in Fig.~\ref{fig:mixing_energies} show more pronounced asymmetry (except Hf$_{1-x}$Zr$_x$N and Zr$_{1-x}$Ti$_x$N) with the $x\approx1$ side having larger positive mixing energies.
The reason for these different trends could be related to differences in the SQS supercells between our study and Ref.~\cite{Holec2013}.
}

We notice that all alloys, except HfN and Al-rich alloys, prefer the planar B$_k$ structure.
We interpret the tendency to relax toward the planar B$_k$ structure ($u$ parameter $\sim0.5$) as an effect induced by the dynamical instability (imaginary phonon frequencies \cite{Isaev2007}, see Fig. 3g in Ref.~\cite{arango-ramirezImportanceHexagonalPhases2020}) combined with an energetic preference for the B$_k$ hexagonal polymorph \cite{holecPhaseStabilityAlloyrelated2011}.
Additionally, the B4 ZrN and HfN binaries are predicted to be dynamically stable in the B$_k$ phase \cite{arango-ramirezImportanceHexagonalPhases2020}.

Previous \emph{ab initio} results \cite{holecPhaseStabilityAlloyrelated2011} and Fig.~\ref{fig:mixing_energies} suggest that $X_{1-x}$Al$_x$N ($X=$ Ti, Zr, Hf) solid solutions energetically favor the B1 structure for Al contents $x\lesssim0.5$, whereas the wurtzite B4 alloy phase becomes the most stable for high $x$ ($\gtrsim0.7$).
At ambient conditions, the ordered TiN, HfN, and ZrN favor the B1 structure, which is also their ground state.
Other recent \emph{ab initio} results indicate that TiN can be metastable in the B4 structure (see Fig. 3g in Ref.~\cite{arango-ramirezImportanceHexagonalPhases2020}).
\red{Our calculations, however, do not find TiN to be stable in the B4 structure}.
Note, indeed, that Ref.~\cite{arango-ramirezImportanceHexagonalPhases2020} reports phonon dispersion results only along a few high-symmetry directions.
Evaluation of the B4 TiN phonon-density of states may be necessary to reveal imaginary frequencies.

\subsection{Machine learning results} \label{subsec:ml_results}
\begin{table}
	\caption{\label{tab:MAE_table_cgcnn} The MAE (in GPa) and MARE (in $\%$) for the test set and the disordered nitrides set.
	The B1$+$B3 means the combination of the B1 and B3 structure sets. 
	}
	\begin{tabular}{lrr}
		\hline
		\hline
		\text{MAE} & $B$ & $G$\\
		\hline
		\text{test set} & 11.6 & 9.0\\
		&&\\
		\text{\underline{disordered nitrides}} & & \\
		\text{unrelaxed B1}  & 15.8 & 28.3\\
		\text{unrelaxed B3}  &  3.9 & 23.6\\
		\text{unrelaxed B1+B3} &  9.7 & 25.9\\
		&&\\
		\text{relaxed B1}  & 14.0 & 23.9\\
		\text{relaxed B3}  &  5.3 &  7.9\\
		\text{relaxed B1+B3} & 9.5 & 15.7\\       
		\hline
		\hline
		&&\\
		\hline
		\hline
		\rowcolor{red}\text{MARE} & $B$ & $G$\\
		\hline
		\rowcolor{red}\text{test set} & 18.4 & 21.7\\
		&&\\
		\rowcolor{red}\text{\underline{disordered nitrides}} & & \\
		\rowcolor{red}\text{unrelaxed B1}  & 6.0 & 16.4\\
		\rowcolor{red}\text{unrelaxed B3}  & 2.1 & 23.7\\
		\rowcolor{red}\text{unrelaxed B1+B3} & 4.0 & 20.2\\
		&&\\
		\rowcolor{red}\text{relaxed B1}  & 5.3 & 13.9\\
		\rowcolor{red}\text{relaxed B3}  & 2.9 & 9.8\\
		\rowcolor{red}\text{relaxed B1+B3} & 4.1 & 11.8\\       
		\hline
		\hline
	\end{tabular}
\end{table}
\begin{figure}
	\begin{center}
		\includegraphics[keepaspectratio,width=1.0\linewidth,height=0.8\textheight]{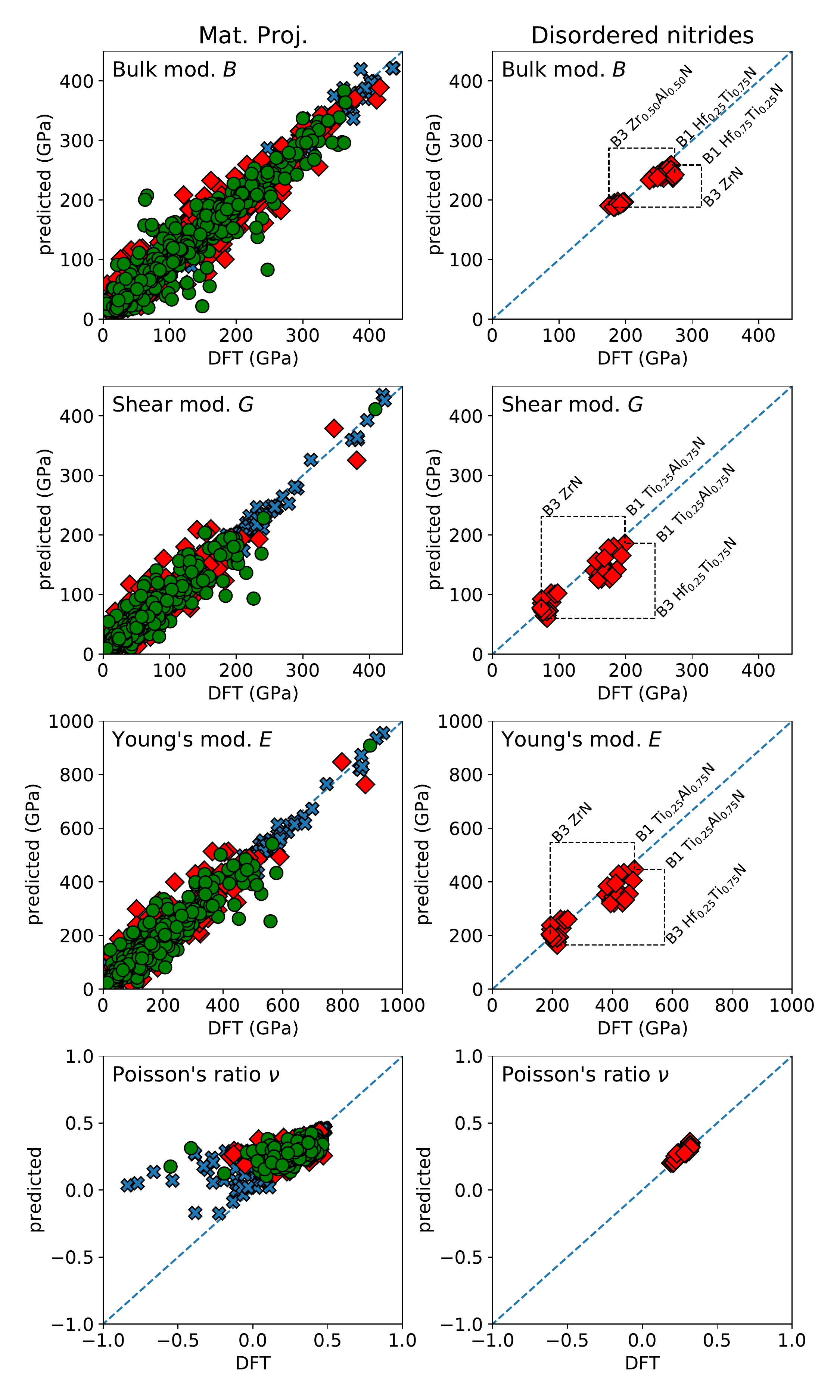}
	\end{center}
	\caption{\label{fig:parity_plot}
		Parity plot of the DFT and ML predicted results. The dashed diagonal line indicates perfect agreement.
		In the Mat. Proj. column blue cross markers indicate training set, red diamonds validation set, and green disks test set.
		The minimum and maximum DFT and ML prediction values for the disordered nitrides are also labeled.}
\end{figure}
Given that it is relatively resource heavy to calculate the elastic tensor of disordered structures, a large amount of supercomputer time would be needed to amass a sizable database of such materials.
We are therefore interested in different strategies to supplement or augment the database using ML techniques.
One way to tackle the problem of low data availability is to train the ML model to a part of the available data that is related to the problem at hand, and then appropriately modifying and verifying the model with the smaller dataset.
This is the basic idea behind TL \cite{Lee2021}.
In this work we use an approach that might be considered a zeroth-order approximation to TL.
In this approach we train an ML model using Materials Project data, which are ordered compounds, and see how well that model predicts the disordered alloys simulated in the present study.
In other words, we check how well the patterns learned from the ordered dataset, without modifications, transfer or generalize to the disordered dataset.
If the ML model generalizes well, one can employ the network to predict, or at least estimate the properties of disordered systems outright, as the case may be, or use it as a basis for further training with the disordered dataset \cite{Yamada2019}.
Performing more sophisticated TL, such as fine-tuning and layer freezing \cite{Yamada2019,Lee2021}, is out of the scope of this study and will be addressed in future work.

\red{Throughout the ML section, the B4 phase is not included because most of the studied B4 alloys turns out to be unstable and transform into Bk structure upon structural relaxation. At the same time, information on compounds with Bk structure in the databases is insufficient to build reliable ML models: hexagonal systems (and the Bk in particular) are underrepresented in the training set (only $\sim1000$ compounds are hexagonal out of the total ~ 8000). One may view the inability to describe B4/Bk alloys as a limitation of the applicability of the ML and TL that requires further in-depth study.
Additionally, those ordered binary compounds (concentration parameter $x = 0$ or $1$) that can be found in the Materials Project data are excluded, because for those data points the error would be biased as they were part of the training set.}

The ML architecture that we use is the CGCNN, and Mazhnik \emph{et al.} have shown that it can be succesfully harnessed, together with Materials Project data, to predict simultaneously both the bulk and shear moduli of an input crystal structure.
One could always train separate ML models for the bulk and shear moduli, but in the approach of Mazhnik \emph{et al.} both quantities are treated on equal footing.
Based on their good results, we are encouraged to implement a similar approach here.
We employ a typical train-validate-test workflow to build the optimal CGCNN model from the Materials Project data.

The performance of the finished model is depicted in Table~\ref{tab:MAE_table_cgcnn} in terms of MAE and MARE for the test set and the disordered dataset, with and without atomic relaxations.
\red{In the Table, the B1$+$B3 means the combination of the B1 and B3 structure sets.}
Table~\ref{tab:MAE_table_cgcnn} shows that the predictions made by the ML model for the disordered alloys with B1 and B3 crystal structure are quite accurate, even though no actual TL was done to the model.
We notice that for the B$3$ structures the shear modulus prediction error is noticeably better for the relaxed data compared to unrelaxed data.
This difference reflects the differences in unrelaxed and relaxed shear moduli in Table~\ref{tab:MAE_table_ar}.
There are two possible reasons that could explain this behavior.
Firstly, the ML prediction error is smaller for the relaxed data, which makes sense since the model was trained on Materials Project data, for which atomic relaxation is used systematically.
It is reasonable to assume that the bonding lengths of the unrelaxed structures appear unphysical to the ML model.
That would explain why the model performs poorly for unrelaxed structures.
Secondly, the CGCNN construction includes a certain cutoff parameter ($\SI{8}{\angstrom}$ in this work) that is used to make the descriptor of each atomic site finite; neighboring atomic sites that fall outside the cutoff radius are neglected.
When the structure is relaxed, some atomic sites may move beyond the cutoff radius while others may move within the radius.
The descriptors between relaxed and unrelaxed structures are therefore slightly different, which could create discrepancies in the ML results for relaxed and unrelaxed structures.

An interesting observation is that the test set and disordered nitrides MAEs show opposite qualitative trends; for the test set, the shear modulus MAE is smaller than the bulk modulus MAE, while for the disordered data this trend is reversed.
Overall, the average ML prediction error for the disordered nitrides is good for the bulk modulus and somewhat worse for the shear modulus.
We can say that the CGCNN architecture generalizes quite well from ordered systems to disordered ones even without explicit fine-tuning of the ML model.

To get a clearer picture of the performance of the ML model, Fig.~\ref{fig:parity_plot} shows a parity plot of the DFT versus ML predicted values (B4/B$_k$ not included).
The left side of Fig.~\ref{fig:parity_plot} shows the performance of the network for the test set (green round markers), as well as for the training set (blue cross markers).
The figure also shows ML predictions for Young's modulus $E$ and Poisson's ratio $\nu$, which are not directly produced by the ML model, but can be calculated from the predicted bulk and shear moduli.
Our results for the test set are similar to those of Mazhnik \emph{et al}. \cite{Mazhnik2020}.
The prediction accuracy for the disordered nitrides falls within the accuracy range of the Materials Project test set.

The right column of the figure shows how well the ML model with the lowest loss function performs for the disordered nitrides.
We can see a fairly tight clustering around the diagonal and an absense of major outliers.
Clear outliers or significant scatter for the nitrides is not expected anyway, because the disordered dataset is very homogeneous (in terms of the variety of structures and chemical elements) compared to the Materials Project training set, so the ML model should work similarly for all the data points the disordered data.
We can identify some structure specific trends in the ML prediction accuracy.
While the bulk and shear moduli of the B$3$ structures are especially well predicted, the B$1$ shear moduli show a lower accuracy.
For the B1 phase the shear modulus is consistently underestimated and the largest single underestimation is $\approx50$ GPa.
It is not easy to give a simple reason for the inconsistent shear modulus prediction performance, but we can note that the shear modulus, as evidenced by Table~\ref{tab:MAE_table_ar}, is a more sensitive quantity than the bulk modulus.
It is the aim of future studies to see how much fine-tuning the current ML network will improve the prediction accuracy.

\section{Discussion} \label{sec:conclusions}
We have developed an automated high-throughput workflow for building a computational database of disordered hard-coating materials and presented results of our calculations.
The reliability of our data is verified by comparing to the Materials Project data and existing literature.
Moreover, we have trained the CGCNN ML model on ordered compounds from the Materials Project and demonstrated that this model is able to readily predict the bulk and shear moduli of disordered nitrides with sufficient accuracy.
The CGCNN architecture seems to be able to learn such fundamental patterns in the data that these patterns hold regardless of the degree of order, indicating good generalizability of CGCNN from ordered to disordered systems.
Prediction accuracy for disordered systems can be further improved by using the ML model trained on ordered data as the starting point for TL.
Our findings open new ways to gain insight into disordered hard coating materials, as well as to support and possibly speed up the investigations of disordered hard coating materials, which are computationally demanding to calculate and thus slow to accumulate into a large database.

\section{Methods} \label{sec:methods}
\subsection{DFT calculations} \label{subsec:dft_methods}
The DFT calculations were performed using the Vienna Ab initio Simulation Package (VASP) \cite{Kresse1996a,Kresse1996b}.
The exchange and correlation effects were treated at the generalized gradient approximation (GGA) level \cite{Langreth1983,Perdew1985,Perdew1986} using the Perdew-Burke-Ernzerhof (PBE) \cite{Perdew1996a} functional.
\red{Energy cutoff is set to \SI{500}{\electronvolt} in the preliminary relaxation phase and to \SI{700}{\electronvolt} in subsequent phases.
The precision mode is set to Accurate (PREC=Accurate).
The Gaussian smearing scheme with a smearing width of \SI{0.05}{\electronvolt} is used.
Ionic relaxations are stopped when all forces are below the threshold $\SI{0.01}{\electronvolt\per\angstrom}$.
In cases where the threshold $\SI{0.01}{\electronvolt\per\angstrom}$ is difficult to reach, the threshold is increased to $\SI{0.03}{\electronvolt\per\angstrom}$ or $\SI{0.05}{\electronvolt\per\angstrom}$.
The elastic constants calculations include the support grid for augmentation charges (ADDGRID=.TRUE.).}

The VASP calculations were managed with The High-Throughput Toolkit (\emph{httk}) \cite{httk_website,Armiento2020}.
The \emph{httk} software offers workflows to create input files, manage calculations on computing clusters, automatically fix broken calculations \red{by adjusting VASP input settings}, and organize results in databases.
The automatic computation of elastic constants is implemented in \emph{httk}.

\begin{figure}
	\begin{center}
		\includegraphics[keepaspectratio,width=1.0\linewidth,height=0.8\textheight]{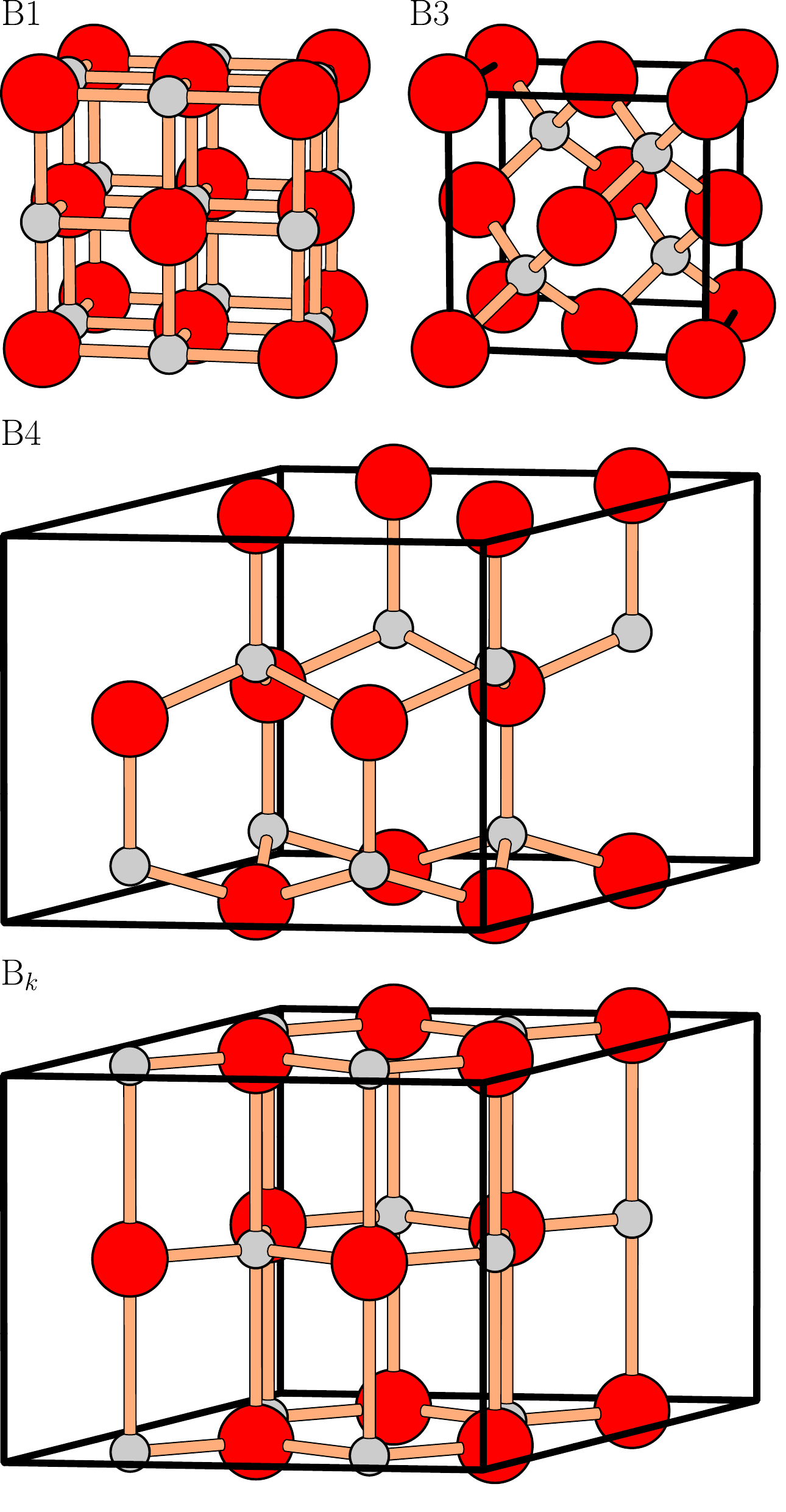}
	\end{center}
	\caption{\label{figure3} The different structures considered in this work.
		The larger red spheres indicate metallic sublattice and the smaller gray spheres nitrogen sublattice.}
\end{figure}

In this work we consider disordered binary and ternary nitrides in cubic rocksalt (B1) \cite{B1_prototype}, cubic zincblende (B3) \cite{B3_prototype}, and hexagonal wurtzite (B4) \cite{B4_prototype} structures.
These structure types are illustrated in Fig. \ref{figure3}.
Disorder is modeled using the special quasirandom structures (SQS) technique \cite{Zunger1990,Wei1990a}.
\red{The SQS supercell that are used in this study are listed in Supplementary materials.}
Here, substitutional disorder is considered only on the metallic sublattice, i.e., the nitrogen sublattice is ordered because it is fully occupied by nitrogen atoms.
All SQS supercells had size of 96 atoms.
This size has been found to be within optimal range for ternary nitrides from the standpoint of accuracy versus computational speed \cite{Tasnadi2012}.
The B1 $X_{0.5}Y_{0.5}$N SQS cell is taken from the supplementary materials of Ref.~\cite{Tasnadi2012}, where it is referred to as the ``$(4\times4\times3)$'' cell.
In Ref.~\cite{Tasnadi2012} the ``$(4\times4\times3)$'' cell was found to model disorder at a high-quality level, that is, comparable to SQS cells of bigger sizes.
The B3 $X_{0.5}Y_{0.5}$N SQS can be easily derived from the B1 $X_{0.5}Y_{0.5}$N SQS by noting that the B1 conventional unit cell turns into B3 when the N atom is shifted from the high-symmetry position $\frac{1}{2}\vec{a}_1 + \frac{1}{2}\vec{a}_2 + \frac{1}{2}\vec{a}_3$ to another high-symmetry position $\frac{1}{4}\vec{a}_1 + \frac{1}{4}\vec{a}_2 + \frac{1}{4}\vec{a}_3$.
All the other SQS supercells are generated with the stochastic Monte-Carlo SQS program mcsqs \cite{VanDeWalle2013}, which is part of the Alloy Theoretic Automated Toolkit (ATAT) program package \cite{VandeWalle2002b}.

Although the SQS scheme is an efficient way to simulate disorder, the downside is that the proper (macroscopic) symmetry is broken and the generated SQS cells typically only have triclinic symmetry.
For the calculation of elastic constants this poses a problem, because we are interested in elastic constant tensors of cubic (B1 and B3) or hexagonal (B4) classes instead of triclinic-structure elastic tensors.
In order to derive properly symmetrized elastic constants from the full triclinic elastic tensor, we employ the projection technique that has been discussed in Ref.~\cite{Tasnadi2012} and references therein.
\red{The ``unprojected'' triclinic elastic constants are reported in the Supplementary materials.
We can see that the deviations between the different crystallographic directions are small, even though the disordered SQS supercells do not respect the proper macroscopic symmetry.
As Ref.~\cite{Tasnadi2012} shows, the projected elastic constants have the desirable property that they converge fast to the ``correct'' value as a function of supercell size, faster than the unprojected elastic constants.}
To facilitate automated calculations, we have implemented the symmetrization technique in \emph{httk}.
The elastic constants in VASP are calculated using the stress-strain method in the same general way that was used, e.g., in Ref.~\cite{Jong2015}.
\red{Four distorted structures are generated for each strain component (in terms of Voigt notation) and the range of distortion are $-3\%$, $-1.5\%$, $1.5\%$, $3\%$ for non-shear components and $-6\%$, $-3\%$, $3\%$, $6\%$ for shear components.}
All polycrystalline quantities are calculated using the Hill averaging scheme \cite{Jong2015}.
The mechanical stability is checked based on the Born-Huang stability criteria \cite{mouhatNecessarySufficientElastic2014}.
In addition to the normal stability check, we also use the stricter tolerance based check of Ref.~\cite{Jong2015}.
For example, cubic systems must fulfill a condition $C_{11} > |C_{12}|$ to be mechanically stable.
In the tolerance based check, a more rigorous condition $C_{11} > \epsilon|C_{12}|$ ($\epsilon>1$) must be fulfilled.
All of the calculated alloys are found to be mechanically stable.
Dynamical stability checks, which are based on phonon dispersion calculations, are out of the scope of this work, but based on existing literature some of the calculated alloys can be expected to be dynamically unstable.

\subsection{Machine learning} \label{subsec:ml_methods}
The ML part of this work is done using the PyTorch package.
The CGCNN \cite{Xie2018} architecture is applied using the code available at GitHub as the basis \cite{cgcnn_github}.
\red{In the CGCNN model the input crystal structure is transformed into a crystal graph, which is constructed from node feature vectors $\bm{v}_i$ and edge feature vectors $\bm{u}_{ij}$.
The node feature vector encodes information about the type of atom located at atomic site $i$.
The atomic information is encoded in an integer vector using one-hot or dummy encoding.
For example, if the training data includes elements N, Al, and Ti, then the elements can be encoded as $\text{N}\Rightarrow[1,0,0]$, $\text{Al}\Rightarrow[0,1,0]$, and $\text{Ti}\Rightarrow[0,0,1]$.
The edge feature vector encodes information about the bonds that the atom at site $i$ makes with its nearest neighbours.
This bonding information is encoded by discretizing a Gaussian distribution function centered around the nearest neighbours of the reference site $i$.
The feature vector for site $i$ with one of its nearest neighbours $j$ is then calculated as
\begin{equation}
\bm{u}_{ij} = \exp{\left(-\frac{(d_{ij}-\bm{\alpha})^2}{h^2}\right)},
\end{equation}
where $d_{ij}$ is the distance between sites $i$ and $j$, $h$ is a discretization step size, and $\alpha$ is a vector of discretized trial distances
\begin{equation}
\bm{\alpha} = \left[ 0, h, 2h, 3h, \dots, R_{\rm cut} \right],
\end{equation}
where $R_{\rm cut}$ is a cutoff radius to make the vector finite in length.
We can see that when $\bm{\alpha}$ is close to $d_{ij}$, $\bm{u}_{ij}$ has values close to one, and when $\bm{\alpha}$ deviates from $d_{ij}$, $\bm{u}_{ij}$ is close to zero.
The crystal graph is then fed into a convolutional neural network, which consists of convolutional layers and pooling layers.
The convolutional layers iteratively modify the node feature vector of site $i$ by convolving it with the edge feature vector $\bm{u}_{ij}$ and node feature vectors of surrounding sites $j$.
The convolution is described by the equation}

\begin{figure}
	\begin{center}
		\includegraphics[keepaspectratio,width=1.0\linewidth,height=0.8\textheight]{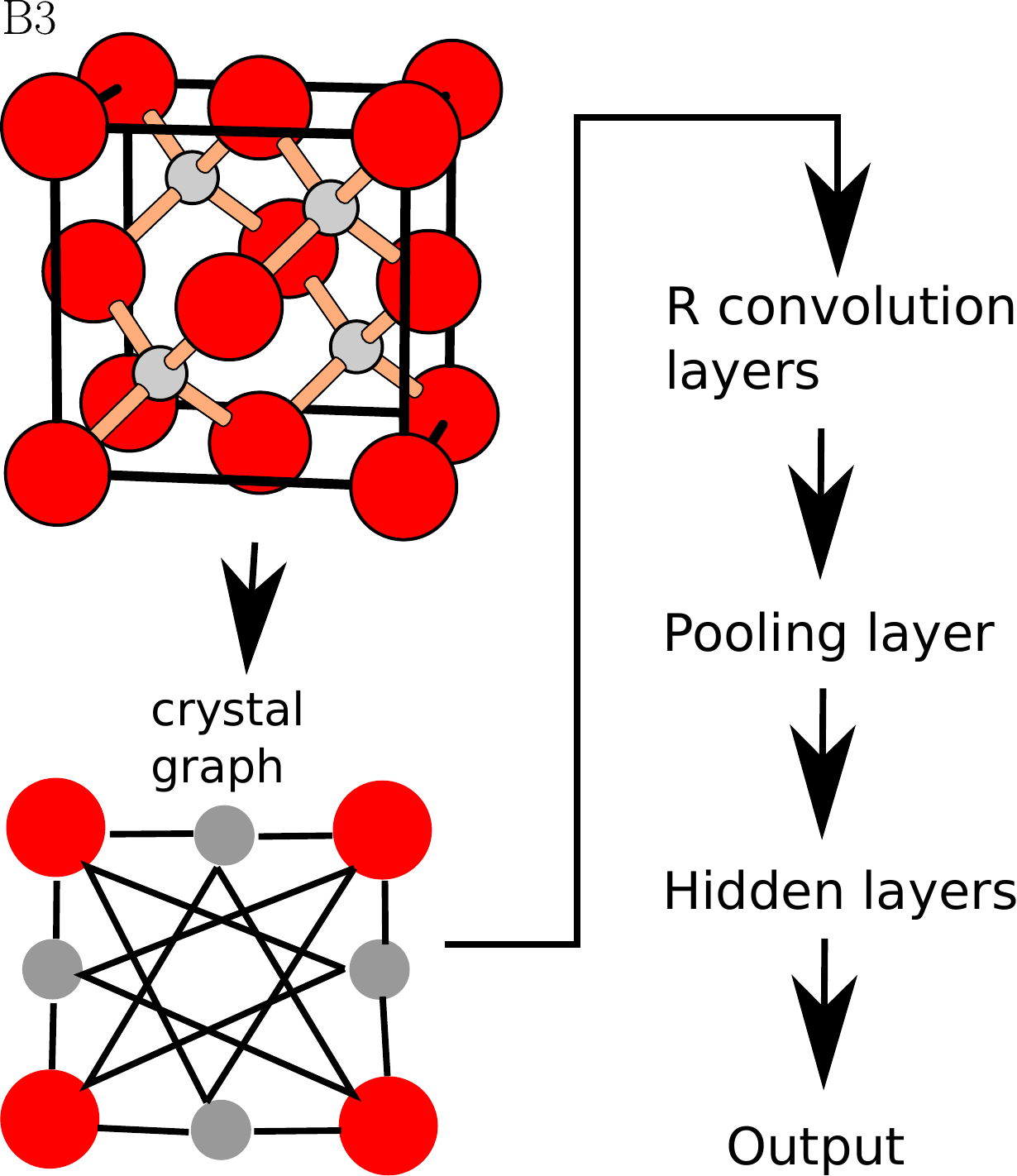}
	\end{center}
	\caption{\label{fig:figure4} Diagram of the CGCNN architecture.}
\end{figure}

\red{
\begin{align}
\bm{v}_i^{(t+1)} = \bm{v}_i^{(t)} &+ \sum_{j}\sigma\left( \bm{z}_{i,j}^{(t)}\bm{W}_f^{(t)} + \bm{b}_f^{(t)} \right)\nonumber\\
&\odot g\left( \bm{z}_{i,j}^{(t)}\bm{W}_s^{(t)} + \bm{b}_s^{(t)} \right),
\end{align}
where $\bm{z}_{i,j}^{(t)} = \bm{v}_i^{(t)} \oplus \bm{v}_j^{(t)} \oplus \bm{u}_{i,j}$ is the concatenation of the node and edge feature vectors, $\sigma$ is the sigmoid activation function, $g$ is the softplus activation function, and $\bm{W}_s^{(t)}, \bm{W}_f^{(t)}, \bm{b}_s^{(t)}, \bm{b}_f^{(t)}$ are the weights and biases of the \emph{t}th convolutional layer~\cite{Xie2018}.
After $R$ convolutional layers (three in this work), a pooling layer is applied to produce the overall feature vector of the crystal $\bm{v}_c$.
The pooling layer is implemented as a normalized summation of all the feature vectors $\bm{v}_1^{(0)}, \bm{v}_2^{(0)}, \dots, \bm{v}_1^{(R)}, \bm{v}_2^{(R)}, \dots, \bm{v}_N^{(R)}$.
Finally, $\bm{v}_c$ is inputed to fully connected hidden layers (two in this work) that predict the output.
A diagram of the CGCNN architecture is presented in Fig.\ref{fig:figure4}.
}

Following Ref.~\cite{Mazhnik2020}, the network is trained to predict bulk and shear moduli using Materials Project data as the training and test datasets.
At the time of writing, Materials Project contains a little over 8000 systems that have elasticity data available.
We randomly split the data into training, validation, and test sets in a 8:1:1 ratio.
\red{While the training set is used to optimize the model parameters, the difference between the validation and test has to do with avoiding biases in the final estimation of the accuracy of the ML model.
Here validation set is used to minimize overfitting in a way that is described below.
This means that the validation set is part of the model optimization process and the model accuracy for the validation set is slightly biased.
Hence we need a test set that only contains data that the final ML model has never seen before.
By evaluating model performance on the test set is therefore a way to minimize biases and maximize the trustworthiness of the model accuracy estimate.}

Before training, the target values (bulk and shear moduli) are normalized using the formula
\begin{equation}
	\widetilde{\bm{y}} = \frac{\bm{y} - \bar{\bm{y}}}{\sigma(\bm{y})},
\end{equation}
where $\bar{\bm{y}}$ is the mean and $\sigma(\bm{y})$ is the standard deviation of the target vector $\bm{y}$.
The loss function to be minimized is defined in terms of the normalized bulk modulus $\widetilde{\bm{B}}$ and shear modulus $\widetilde{\bm{G}}$ as
\begin{equation}
	\mathcal{L} = \frac{\sum_{i} (\widetilde{B}_i^{\rm pred} - \widetilde{B}_i^{\rm DFT})^2 + (\widetilde{G}_i^{\rm pred} - \widetilde{G}_i^{\rm DFT})^2}{N},
\end{equation}
where $N$ is the number of data points.
Based on the information of Refs.~\cite{Xie2018,Mazhnik2020,Lee2021} and their supplementary materials, as well as our own testing, we can infer reasonable values for the hyperparameters without performing extensive hyperparameter tuning.
The crystal structure descriptor related settings are the same as those reported in Ref.~\cite{Mazhnik2020}.
The training process uses the Adam optimizer with a learning rate of 0.005 and weight decay 0.0.
Training is continued for a minimum of 500 epochs to give the randomly initialized model parameters enough time to develop.
Training is stopped after 1500 epochs, as it is unlikely that any significant progress will happen after that point.
In order to avoid the training process getting stuck in a suboptimal local minimum, 30 instances were trained.
To reduce random variance, an ensemble of five networks with the lowest validation loss is used to make predictions.
The validation set is used to decide the final model parameters by extracting the parameters from the epoch that yields the loss function minimum for the validation set.
This is one form of the early stopping technique and the purpose is to mitigate overfitting issues \cite{Hastie2009}.
Finally, an unbiased estimate of the performance of the finished model is obtained by evaluating the model using the unseen test set (see Table~\ref{tab:MAE_table_cgcnn}).
For further details about the CGCNN architecture and the training process, the reader is referred to Refs.~\cite{Xie2018,Mazhnik2020,Lee2021}.

\section{Data availability} \label{sec:data_availability}
Relevant calculated DFT data and instructions to download the Materials Project data, as well as the trained ML model to reproduce the results of this paper are provided as Supplementary materials.

\section{Code availability} \label{sec:code_availability}
The code to reproduce the results of this paper are provided as Supplementary materials.

\section{Acknowledgments}
We gratefully acknowledge financial support from the Competence Center Functional Nanoscale Materials (FunMat-II) (Vinnova Grant No. 2016–05156).
Support from the Knut and Alice Wallenberg Foundation (Wallenberg Scholar Grant No. KAW-2018.0194), the Swedish Government Strategic Research Areas in Materials Science on Functional Materials at Linköping University (Faculty Grant SFO-Mat-LiU No. 2009 00971) and SeRC is gratefully acknowledged.
Theoretical analysis of results of first-principles calculations was supported by the Russian Science Foundation (Project No. 18-12-00492).
R.A. acknowledges support from the Swedish Research Council (VR) Grant No. 2020-05402 and the Swedish e-Science Centre (SeRC).
The computations were enabled by resources provided by the Swedish National Infrastructure for Computing (SNIC), partially funded by the Swedish Research Council through grant agreement no. 2018-05973.
The computational resources provided by the PDC Center for High Performance Computing are also acknowledged.

\section{Competing interests}
The authors declare no competing interests.

\section{Author contributions}
H.L., R.A, and I.A.A. conceived the project.
H.L. performed the calculations and analyzed the results.
H.L., F.T., D.G.S., L.J.S.J., R.A., and I.A.A. discussed the results and wrote the manuscript.


%

\ifarXiv
\onecolumngrid
\clearpage
\twocolumngrid
\foreach \x in {1,...,\numbersupplementpages}
{
	\clearpage
	\includepdf[pages={\x,{}}]{\supplementfilename}
}
\fi

\end{document}